\input harvmac
\input epsf
\def\Title#1#2{\rightline{#1}\ifx\answ\bigans\nopagenumbers\pageno0\vskip1in
\else\pageno1\vskip.8in\fi \centerline{\titlefont #2}\vskip .5in}
%

%
%
\ifx\epsfbox\UnDeFiNeD\message{(NO epsf.tex, FIGURES WILL BE IGNORED)}
\def\figin#1{\vskip2in}
\else\message{(FIGURES WILL BE INCLUDED)}\def\figin#1{#1}
\fi
\def\Fig#1{Fig.~\the\figno\xdef#1{Fig.~\the\figno}\global\advance\figno
 by1}
%
%
%
%
\def\ifig#1#2#3#4{
\goodbreak\midinsert
\figin{\centerline{\epsfysize=#4truein\epsfbox{#3}}}
\narrower\narrower\noindent{\footnotefont
{\bf #1:}  #2\par}
\endinsert
}

%
%
\font\ticp=cmcsc10
\def\sq{{\vbox {\hrule height 0.6pt\hbox{\vrule width 0.6pt\hskip 3pt
   \vbox{\vskip 6pt}\hskip 3pt \vrule width 0.6pt}\hrule height 0.6pt}}}

\def\ajou#1&#2(#3){\ \sl#1\bf#2\rm(19#3)}
\def\jou#1&#2(#3){,\ \sl#1\bf#2\rm(19#3)}
\def\hf{{1\over 2}}

\def\frac#1#2{{#1\over#2}}
\def\Rads{{R}}

\def\eg{{\it e.g.}}
\def\etc{{\it etc.}}

\def\caln{{\cal N}}
\def\ptil{{\tilde p}}

\def\calo{{\cal O}}

%
%
\lref\Suss{L. Susskind, private communication.}
\lref\Levin{
E.~Levin,
``Everything about Reggeons. I: Reggeons in *soft* interaction,''
arXiv:hep-ph/9710546.
}
\lref\GuKP{
S.~S.~Gubser, I.~R.~Klebanov and A.~W.~Peet,
``Entropy and Temperature of Black 3-Branes,''
Phys.\ Rev.\ D {\bf 54}, 3915 (1996)
[arXiv:hep-th/9602135].
}
\lref\Heis{W. Heisenberg, Z.\ Phys.\ {\bf 133}, 65 (1952).}
\lref\Mald{
J.~Maldacena,
``The large $N$ limit of superconformal field theories and supergravity,''
Adv.\ Theor.\ Math.\ Phys.\  {\bf 2}, 231 (1998)
[Int.\ J.\ Theor.\ Phys.\  {\bf 38}, 1113 (1998)]
[arXiv:hep-th/9711200].
}
\lref\FSS{
S.~B.~Giddings,
``Flat-space scattering and bulk locality in the AdS/CFT  correspondence,''
Phys.\ Rev.\ D {\bf 61}, 106008 (2000)
[arXiv:hep-th/9907129].
}
\lref\PoSt{
J.~Polchinski and M.~J.~Strassler,
``Hard scattering and gauge/string duality,''
Phys.\ Rev.\ Lett.\  {\bf 88}, 031601 (2002)
[arXiv:hep-th/0109174].
}
\lref\GiTh{
S.~B.~Giddings and S.~Thomas,
``High energy colliders as black hole factories: 
The end of short  distance physics,''
Phys.\ Rev.\ D {\bf 65}, 056010 (2002)
[arXiv:hep-ph/0106219].
}
\lref\DiLa{
S.~Dimopoulos and G.~Landsberg,
``Black holes at the LHC,''
Phys.\ Rev.\ Lett.\  {\bf 87}, 161602 (2001)
[arXiv:hep-ph/0106295].
}
\lref\BaFi{
T.~Banks and W.~Fischler,
``A model for high energy scattering in quantum gravity,''
arXiv:hep-th/9906038.
}
\lref\EaGi{
D.~M.~Eardley and S.~B.~Giddings,
``Classical black hole production in high-energy collisions,''
arXiv:gr-qc/0201034.
}
\lref\GoWi{
W.~D.~Goldberger and M.~B.~Wise,
``Modulus stabilization with bulk fields,''
Phys.\ Rev.\ Lett.\  {\bf 83}, 4922 (1999)
[arXiv:hep-ph/9907447].
}
\lref\HoPo{
G.~T.~Horowitz and J.~Polchinski,
``A correspondence principle for black holes and strings,''
Phys.\ Rev.\ D {\bf 55}, 6189 (1997)
[arXiv:hep-th/9612146].
}
\lref\RaSui{
L.~Randall and R.~Sundrum,
``A large mass hierarchy from a small extra dimension,''
Phys.\ Rev.\ Lett.\  {\bf 83}, 3370 (1999)
[arXiv:hep-ph/9905221].
}
\lref\GaTa{
J.~Garriga and T.~Tanaka,
``Gravity in the brane-world,''
Phys.\ Rev.\ Lett.\  {\bf 84}, 2778 (2000)
[arXiv:hep-th/9911055].
}
\lref\GKR{
S.~B.~Giddings, E.~Katz and L.~Randall,
``Linearized gravity in brane backgrounds,''
JHEP {\bf 0003}, 023 (2000)
[arXiv:hep-th/0002091].
}
\lref\PoSti{
J.~Polchinski and M.~J.~Strassler,
``The string dual of a confining four-dimensional gauge theory,''
arXiv:hep-th/0003136.
}
\lref\CGRT{
C.~Csaki, M.~Graesser, L.~Randall and J.~Terning,
``Cosmology of brane models with radion stabilization,''
Phys.\ Rev.\ D {\bf 62}, 045015 (2000)
[arXiv:hep-ph/9911406].
}
\lref\GKP{
S.~B.~Giddings, S.~Kachru and J.~Polchinski,
``Hierarchies from fluxes in string compactifications,''
arXiv:hep-th/0105097.
}
\lref\Gubs{
S.~S.~Gubser,
``Curvature singularities: The good, the bad, and the naked,''
arXiv:hep-th/0002160.
}
\lref\DFGK{
O.~DeWolfe, D.~Z.~Freedman, S.~S.~Gubser and A.~Karch,
``Modeling the fifth dimension with scalars and gravity,''
Phys.\ Rev.\ D {\bf 62}, 046008 (2000)
[arXiv:hep-th/9909134].
}
\lref\Isra{
W.~Israel,
``Singular Hypersurfaces And Thin Shells In General Relativity,''
Nuovo Cim.\ B {\bf 44S10}, 1 (1966)
[Erratum-ibid.\ B {\bf 48}, 463 (1966)].
}
\lref\APR{
N.~Arkani-Hamed, M.~Porrati and L.~Randall,
``Holography and phenomenology,''
JHEP {\bf 0108}, 017 (2001)
[arXiv:hep-th/0012148].
}
\lref\DiSu{S. Dimopoulos, R. Emparan, and L. Susskind, unpublished.}
\lref\Snow{
S.~B.~Giddings,
``Black hole production in TeV-scale gravity, and the 
future of high  energy physics,''
in {\it Proc. of the APS/DPF/DPB Summer Study on 
the Future of Particle Physics (Snowmass 2001) } ed. R.~Davidson and C.~Quigg,
arXiv:hep-ph/0110127.
}
\lref\JaPe{R.~A.~Janik and R.~Peschanski,
``Reggeon exchange from AdS/CFT,''
Nucl.\ Phys.\ B {\bf 625}, 279 (2002)
[arXiv:hep-th/0110024]; R.~A.~Janik,
``High energy scattering and AdS/CFT,''
Acta Phys.\ Polon.\ B {\bf 32}, 4105 (2001)
[arXiv:hep-th/0111104].
}

\Title{\vbox{\baselineskip12pt
\hbox{hep-th/0203004}\hbox{SU-ITP-02/09}
}}
{\vbox{\centerline{High energy QCD scattering, the shape of gravity}
\vskip2pt\centerline{on an IR brane, and the Froissart bound}
}}
\centerline{{\ticp Steven B. Giddings}\footnote{$^\dagger$}
{Email address:
giddings@physics.ucsb.edu} }
\bigskip\centerline{ {\sl Department of Physics}}
\centerline{\sl University of California}
\centerline{\sl Santa Barbara, CA 93106-9530}
\centerline{and}
\centerline{ {\sl Department of Physics and SLAC}}
\centerline{\sl Stanford University}
\centerline{\sl Stanford, CA 94305/94309}
\bigskip
\bigskip
\centerline{\bf Abstract}
High-energy scattering in non-conformal gauge theories is investigated
using the AdS/CFT dual string/gravity theory.  It is argued that
strong-gravity processes, such as black hole formation, play an important
role in the dual dynamics.  Further information about this dynamics is
found by performing a linearized analysis of gravity for a mass near an
infrared brane; this gives the far field approximation to black hole or
other strong gravity effects, and in particular allows us to estimate
their shape.
From this shape, one can infer a total scattering cross-section that grows
with center of mass energy as $\ln^2E$, saturating the Froissart bound.

\Date{}

\newsec{Introduction}

A dominant theme in the past few years of string theory has been the
conjectured gauge theory/string theory duality of Maldacena\refs{\Mald}.
We are still far from anything approaching a proof of this duality, and
we face serious challenges on the most interesting question of 
using it to infer local properties of
string theory (see \eg\ \refs{\FSS}) but considerable evidence has been
amassed that one can derive features of gauge theory from bulk
gravitational physics.  Much of what was initially learned dealt with
quantities highly constrained by symmetries, such as the mass spectrum, but
more recently attention has been turned to investigating dynamical
properties of the theory that are not as tightly circumscribed.  This has
gone hand in hand with study of spacetimes corresponding to 
gauge theory vacua with less than the maximal ${\cal
N}=4$ symmetry; these are typically found by deforming the lagrangian on
the gauge theory side, and correspondingly turning on non-normalizable
modes of bulk fields on the string theory side.  A
particularly interesting recent example is the work of Polchinski and
Strassler\refs{\PoSt}, which addresses 
a glaring puzzle in the correspondence:  how
is it that string scattering, which is inherently soft at high energies,
reproduces the familiar hard behavior of QCD?   (For other discussions of 
high-energy scattering in QCD via AdS/CFT, see \refs{\JaPe} and 
references therein.)  By analyzing
scattering in the bulk, they find that the soft behavior of string theory
conspires with the shape of the bulk wavefunctions to produce the correct
power-law behavior.  We can now ask what else we might learn about QCD from
the bulk perspective.  One obvious question immediately presents itself:
what can one say about quantities like the total cross section in very high
energy scattering?

Another theme that has recently gained interest is that of the
contribution of strong gravitational effects to high energy scattering.  
For example, in TeV-scale gravity scenarios, based either on large extra
dimensions or on strongly warped spacetimes, black holes should be produced
once scattering energies pass the fundamental Planck scale near a TeV, and
these processes are a very exciting aspect of the phenomenology of these
models\refs{\GiTh,\DiLa} (for a review, see \refs{\Snow}).  
One can readily estimate the relevant
cross-sections based on the observation that at very high energies black
hole formation should be well approximated classically (see \eg\
\refs{\BaFi}).  The na\"\i ve classical estimate of the cross-section to
produce a black hole,
\eqn\crossest{\sigma\sim \pi r_h^2(E)\ ,}
where $r_h$ is the Schwarzschild radius corresponding to center of mass
energy $E$, is a good guide to the magnitude of these effects, as has been
confirmed by a more detailed recent analysis of classical high-energy
scattering\refs{\EaGi}.  Because of it's growth -- like $E^{2/D-3}$ in $D$
flat dimensions -- this cross section is believed to be a dominant feature of
high energy gravitational scattering.  Now a second obvious question
presents itself:  what is the role of strong gravity and in particular
black holes in the dual gauge theory?

Indeed, as this paper will argue, our two questions are intimately
related.  Strong gravitational effects are a dominant feature of the
high-energy gauge theory scattering, and an estimate analogous to
\crossest, using properties of gravity in deformed AdS backgrounds, yields
a high energy cross-section of the form
\eqn\Froiss{\sigma \sim {1\over m^2} \ln^2 \left({E\over E_0}\right)\ ,}
where $m$ is the mass of the lightest excitation.  This behavior saturates
the Froissart bound, which is implied by unitarity and 
is believed by many to describe the correct
high-energy behavior of QCD.

A more detailed description of the resulting gauge-theory physics remains
to be found, but if black hole formation in the bulk is the dominant
process at high energy, we might hope to discover the dual physics of
Hawking radiation and other interesting effects.  At present, however, a
significant obstacle appears from questions, raised by L. Susskind,
about whether the relevant
gravitational solutions are stable.  This is not crucial for them to
produce the cross section \Froiss, but it is critical in determining the
subsequent evolution of the black hole and the corresponding gauge theory
dynamics.  Regardless, it appears that one may be able to think of the
gauge theory physics as corresponding to a ``fireball'' that then decays,
more or less rapidly depending on these stability questions, into an
approximately thermal state.

In outline, this paper will first review the basic setup of a truncated AdS
space corresponding to non-conformal gauge theory, along the lines of
\PoSt.  This is followed by a discussion of the scales relevant to strong
gravitational effects.  At moderately high 
energies one expects to produce black holes
small as compared to the anti-de Sitter radius $R$, but at higher energies
these should grow larger than $R$.  Much of the paper deals with the
problem of determining their shape.
In the full non-linear theory this is a very
difficult problem, but it is argued that a linear analysis can give
substantial information about the shape of such solutions.   The basic
linear equations and boundary conditions are given in section three.
A critical
role is played by the dynamics that stabilizes the infrared end of the
space (the ``radion''), and section four presents a linear analysis of
gravity for both small and large radion masses.  In particular, in the
heavy radion limit solutions are found that correspond to the far-field of
an object that could be a black hole.  In either limit of the radion mass,
one finds cross-section estimates of the form \Froiss.  Section four also
briefly comments on the relevance of this physics to TeV-scale gravity
scenarios in which the extra dimensions may be approximated by a slice of
AdS.  Section five contains further discussion of the dual gauge theory
physics, and section six presents conclusions and several open questions.
There are two appendices:  one that derives the Green function for Anti-de
Sitter space truncated by an infrared boundary -- or brane -- and another
on stabilization mechanisms such as Goldberger-Wise\refs{\GoWi} and their
effective description in geometries whose only boundary is in the infrared.

\newsec{Gauge theory scattering and truncated AdS}

On the gauge-theory side, we are interested in high-energy 
scattering in a large-$N$  $\caln=4$ supersymmetric gauge theory with broken
conformal symmetry and partially broken supersymmetry.  Following
Polchinski and Strassler\refs{\PoSt}, we
will assume that on the gravity-side, this is dual to a supergravity (or
more precisely, superstring) solution with warped
metric 
\eqn\bmet{ds^2= e^{2A(y)} \eta_{\mu\nu} dx^\mu dx^\nu + g_{mn}(y)dy^m dy^n}
that is approximately AdS in a large region, namely 
\eqn\bmett{ds^2\approx {R^2\over z^2}\left(dz^2 + \eta_{\mu\nu} dx^\mu
dx^\nu \right) + R^2 ds_X^2\ }
where $X$ is some appropriate compact manifold.
Here $R$ is the AdS radius, and in terms of gauge theory parameters
satisfies the approximate relation $R^4\sim g^2N \alpha^{\prime 2}$.  
In particular, at long distances we can effectively 
think of the smooth geometry given by
\bmet\ as being truncated in the infrared; in some situations it is
convenient to think of an ``infrared brane'' as lying at this end of the
space.  
Ref.~\refs{\PoSt} take the IR end of the space to lie at
an arbitrary $z$, but by an overall rescaling of coordinates, 
$z\rightarrow \lambda z$,
$x\rightarrow \lambda x$, we can without loss of generality take the space
to end at $z=R$.  This choice also eliminates the need for the different
string scales discussed in \refs{\PoSt}.  Since the scale of the
Kaluza-Klein masses are set by $1/R$ (as we will see in more detail), and
these are interpreted in the gauge theory as glueballs, this means that the
effective QCD scale is $\Lambda_{\rm QCD}\sim 1/R$.

If we perform scattering in this space, the {\it conserved} momentum is the gauge
theory momentum, 
\eqn\cmom{p_\mu = -i {\partial \over \partial x^\mu}\ .}
As reviewed in \PoSt, we can simply relate glueball scattering
amplitudes to bulk scattering as follows.  An incoming glueball state has a
bulk wavefunction of the rough form
\eqn\gluewave{\psi\sim e^{ipx} f(z/R) g(y^i)\ .}
Here $g$ represents modes of the internal manifold $X$, and $f(z/R)$ gives
the radial wavefunction in AdS.  The wavefunctions are exactly of this form
in a space that is $AdS_5\times S^5$ truncated by an infrared boundary at
$z=R$, but in general the underlying smooth physics and deformation from
AdS will lead to additional mixing.  Scattering amplitudes take the form
\eqn\scatamp{{\cal A}_{\rm gauge}(p) = \int d^{10}x \sqrt{-G} {\cal A}_{\rm
bulk}(x^\mu,y^i,z) \prod_i \psi_i \ .}
Details of the exact form of the wavefunctions are not needed for our very
general purposes, but they generically have large support in the vicinity of
the IR boundary.  
The bulk amplitudes depend on the {\it proper} momentum as
measured by a local observer, which for a given conserved momentum is a
function of $z$:
\eqn\pmom{\ptil_\mu(z) = {z\over R} p_\mu\ .}
Note that with our conventions, $\ptil_\mu=p_\mu$ at the IR boundary.

There are several interesting threshholds that we encounter as we consider
scattering at increasing energies.  The first is the threshhold for
scattering in which intermediate string states are important.  This
threshhold is at $p\sim 1/\sqrt{\alpha'}$.  
At higher energies, the proper energy
reaches this scale at a radius given by
\eqn\rscatt{z_{\rm scatt}\sim {R\over\sqrt{\alpha'}p} \ .}
It is this scattering physics that is important for the amplitudes
studied in \PoSt.  However, there are other potentially important effects
that \PoSt\ doesn't include.

Specifically, we know that once proper energies pass the 10d Planck
scale, $M_P \sim g_s^{-1/4}/\sqrt{\alpha'}$, 
gravity becomes strong and in particular black holes
can form.  These may initially be best described as highly excited string
states, but 
by the string
correspondence principle\refs{\HoPo} we expect a smooth crossover from
intermediate string states to intermediate black hole states at the
correspondence scale $E_c\sim g_s^{-2}/\sqrt{\alpha'}$.
This physics is thus relevant for $p\roughly> M_P$.  If the AdS radius (and
other geometrical scales) 
are 
much larger than the Planck length, these black holes can be thought of as
essentially living in flat space.
As we go to higher
energies, two things happen.  First, one can make black holes at smaller
values of $z$, further into the UV region of the geometry.  Any real (as
opposed to virtual) black holes of this form would of course be unstable to
falling towards the boundary.  But perhaps more
importantly, one can make 
larger black
holes.  Indeed, for given energy, the radius of a 10d black hole grows as 
\eqn\rhor{r_h\sim E^{1/7}\ ,}
and consequently the cross-section for black hole production at the IR end
of the space grows as
\eqn\crosss{\sigma\sim E^{2/7}\ .}
In terms of QCD parameters, $M_p\sim N^{1/4}\Lambda_{\rm QCD}$ and $E_c\sim
N^2 \Lambda_{QCD}/(g^2 N)^{7/4}$.  
Above these energies, one expects possibly 
interesting consequences of the black hole
formation for QCD scattering.  One is the power-law growth  \crosss.
Another comes from the energy dependence of the Hawking temperature of a 10d black
hole:
\eqn\hawkt{T_H\sim E^{-1/7}\ .}
As is typical in black hole physics, this corresponds to a negative
specific heat.  Further aspects of this regime have been explored by
Dimopoulos, Emparan, and Susskind\refs{\DiSu}.

As the energy grows further, the black hole size approaches the AdS radius.
This occurs at an energy 
\eqn\largeE{E_R\sim M_P^8 R^7\ .}
In QCD parameters, this corresponds to $E_R\sim N^2 \Lambda_{\rm QCD}$.
At this energy the approximation of these as black holes in background flat
space breaks down.   The precise nature of black holes at larger energies
is an interesting question, but if there are such solutions, we 
expect them to be
completely smeared out over the compact manifold $X$.  Furthermore, effects
of the AdS geometry surrounding the IR brane become important.  Finding the
shape and properties of such a strong gravity region near the IR brane
is a non-trivial problem to which we now turn.

\newsec{Linearized gravity with an IR brane}

We would like to understand properties of gravity in the vicinity of the IR
end of the space \bmet, and at distance scales larger
than $R$.  In particular we would like to determine the nature of possible
large black hole solutions at the IR boundary.
Of course this is in general a very
complicated problem.  
For solutions with sizes $\roughly >R$, we 
expect any such black hole solutions to be
uniformly spread over the compact directions of the internal manifold $X$,
but their shape in the other directions should be non-trivial, and, as
we'll see, their stability is in question.  In order
to simplify the problem, we work in an effective description in which we neglect
$X$ and work about a slice of ordinary AdS space terminated in the infrared at
$z=R$.  Of course, this is just a boundary condition summarizing the much
more complicated underlying smooth 10d geometry.
The dynamics of this slice can be discussed by introducing an
effective tension for an IR brane at $z=R$, as in \refs{\RaSui}, but that
will not be critical to our discussion.  Even with these simplifications,
at present we lack the tools to find non-linear gravitational solutions
corresponding to black holes on branes in AdS.  However, one can
nonetheless find useful information about these solutions from a linearized
analysis of gravity, using the following simple principle:

{\it Linearization principle:}  Suppose that a solution to the full
non-linear field equations exists.  In weak-field regions, there
must exist a corresponding solution of the linearized equations.  One can
therefore infer general properties of the non-linear solution (assuming it
exists) by examining solutions of the linearized equations.

Of course, it may be that there is a linearized solution but no
corresponding non-linear solution, or there may be more than one exact
solution with the same asymptotic properties.  For the present purposes we
assume that such violations of existence and uniqueness don't occur.

Using this method, basic structural properties such as 
shape and size of the solution can be inferred from the shape of the
linearized solutions.  For example, if we are studying a black hole, we can
infer the approximate location and shape of its horizon by finding the
region where the metric perturbation from the background geometry
(Minkowski, AdS, $\cdots$) becomes strong.   The corresponding problem of
finding black holes on a UV brane was treated by a linear analysis in 
\refs{\GaTa,\GKR}, which used this technique to infer the ``pancake'' shaped
black holes of that case.  We next turn to a general discussion of the
linearized equations, their boundary conditions, brane stabilization, and
conditions under which the linearized approximation fails.

\subsec{Linearized gravitational equations}

Our starting point is 
to consider perturbations about the metric \bmett, suppressing $X$; it
is convenient to work using the ``height'' coordinate $y$ given by 
\eqn\yzreln{z=Re^{y/R}\ .}
The perturbed AdS metric takes the form
\eqn\linmet{ds^2 = (1+h_{yy})dy^2 + e^{-2y/R}(\eta_{\mu\nu} + h_{\mu\nu})dx^\mu
dx^\nu\ } 
where we work with a gauge 
\eqn\gaugei{h_{\mu y}=0\ .}
This condition does not completely fix the gauge, and in particular allows
gauge transformations
\eqn\gaugeg{\eqalign{y&\rightarrow y+ \alpha^y\cr x^\mu&\rightarrow x^\mu +
\alpha^\mu}}
with
\eqn\gaugecond{\partial_\mu\alpha^y + e^{-2y/R} \partial_y \alpha_\mu =0\
.}
Here and in the remainder of the paper we adopt the convention that indices
on small d-dimensional quantities ($h_{\mu\nu}$, $\alpha^\mu$, {\it etc.})
are raised and lowered with the metric $\eta_{\mu\nu}$.  The resulting gauge
redundancy is parametrized by functions $\alpha^y(x,y)$ and
$\beta^\mu(x)$.  

There are furthermore two possibilities for allowed $\alpha^y$.  We may
fix the region in which we are working, $y\in (-\infty,0)$, and thus demand
that gauge transformations not translate the boundary,
\eqn\rigidg{\alpha^y(0)=0\ .}
This gauge, which we refer to as ``straight gauge,'' allows us to transform
$h_{yy}$ to depend only on $x$.  If, on the other hand, we allow gauge
transformations that move the boundary, 
\eqn\bentxm{\alpha^y(0) = L(x)\ ,}
for some
function $L$, then we may completely eliminate $h_{yy}(x)$.  In this gauge,
referred to as
``bent gauge,''  the boundary lies at $y=L(x)$.  The four-dimensional fields
$h_{yy}(x)$ or $L(x)$ are 
the present incarnation of the radion field familiar from
two-brane scenarios. In either case, the $y$-independent gauge freedom
\eqn\dgauge{x^\mu\rightarrow x^\mu + \beta^\mu(x)}
remains.

Einstein's equations for a perturbation with $h_{yy}=0$ were given in  
\refs{\GKR}; we'll find both straight and bent gauges to be useful so
generalize these to $h_{yy}\neq0$.
Letting the source stress tensor be $T_{MN}$, defining
a modified metric perturbation 
\eqn\hbardef{{\bar
h}_{\mu\nu}= h_{\mu\nu}-\hf h \eta_{\mu\nu}\ ,  }
with $h=\eta^{\mu\nu}
h_{\mu\nu}$,
and working in $d+1$
dimensions, their perturbations can be shown to be
%
%
%
\eqn\gyy{(\partial_d^2 h- \partial^\mu\partial^\nu h_{\mu\nu})e^{2y/R} -
{d-1\over R} \partial_y h -{d(d-1)\over R^2} h_{yy} = {1\over M_P^{d-1}} 
T_y^y\ ,}
\eqn\muysol{
\partial_y\partial^{\nu} \left(h_{\mu\nu}-h\eta_{\mu\nu}\right)-{d-1\over
R}\partial_\mu h_{yy} = {T_{\mu}^y\over M_P^{d-1}}\ ,
}
and
\eqn\munueq{
   \eqalign {
\sq{ \bar{h}_{\mu\nu} }
&= { \eta_{\mu\nu}\over 2 }\, e^{yd/R} \, \partial_y
(e^{-yd/R}\partial_y{h})\cr
&+ e^{2y/\Rads}
(-\eta_{\mu\nu}\partial^{\lambda}\partial^{\sigma}{\bar{h}_{\lambda\sigma}} +
\partial^{\lambda}\partial_{\mu}{\bar{h}}_{\nu\lambda}
+\partial^{\lambda}\partial_{\nu}{\bar{h}}_{\mu\lambda} )\cr
& + e^{2y/R}\left(\partial_d^2 h_{yy}\eta_{\mu\nu} - \partial_\mu
\partial_\nu h_{yy}\right)
+ {d-1\over R} e^{yd/R}\partial_y\left(e^{-yd/R}h_{yy}\right)\eta_{\mu\nu}\cr
&- { e^{2y/R}\over M_P^{d-1}} \, T_{\mu\nu}\  \cr}
}
from the $(yy)$, $(\mu y)$, and $(\mu\nu)$ components of Einstein's
equations, respectively.

\subsec{Boundary conditions and linear approximation}

The linearized Einstein equations \gyy-\munueq\ 
must be supplemented by boundary conditions.  The boundary conditions
in the UV ($z=0$, $y=-\infty$) 
are simply that the perturbation falls off sufficiently rapidly.  
Appropriate boundary conditions at the IR brane are Neumann
boundary conditions,
\eqn\Neubc{n^I\partial_I h_{\mu\nu}{}|_{\partial} =0\ }
where $n^I$ is the unit outward normal to the boundary.
These can be motivated from orbifold boundary conditions 
\eqn\orbbc{n^I\partial_I h_{\mu\nu}|_{-} = - n^I
\partial_I h_{\mu\nu}|_{+}}
about the boundary, or equivalently from the statement that the
gravitational field vanishes at the boundary.  
In
the case where there is a 
source localized on the IR brane,
\eqn\sourceT{T_{\mu\nu}= S_{\mu\nu}(x)\delta(y)\ ; \ T_{yy}=T_{y\mu}=0\
,}
these are modified; integrating \munueq\ over a small neighborhood of the
brane gives the Israel matching conditions\refs{\Isra}.  
Working in straight gauge, where the boundary is at $y=0$, and combining 
the result with
\orbbc,  gives 
\eqn\bcond{
\partial_y(h_{\mu\nu} - \eta_{\mu\nu}h )\vert_{ y=0} - {d-1\over R}h_{yy}(0) 
\eta_{\mu\nu}=  {S_{\mu\nu}(x)\over
2M_P^{d-1} }\ .}
Alternatively, these may be transformed to bent gauge using \bentxm.  The
resulting boundary conditions become
\eqn\bcondb{
\partial_y(h_{\mu\nu} - \eta_{\mu\nu}h )|_{ y=L} 
- 2\partial_\mu\partial_\nu L +2\eta_{\mu\nu}\partial_d^2L
=  {S_{\mu\nu}(x)\over
2M_P^{d-1} }\ .}

If we find solutions to the equations \gyy-\munueq\ with boundary
conditions \bcond\ or \bcondb, we've found a valid approximate solution as
long as the metric perturbation remains small,
\eqn\smallh{h_{IJ}\ll1\ .}
An additional important condition when working in bent gauge is that the
bending remain small (this was used in deriving \bcondb):
\eqn\smallb{\partial_\mu L \ll 1\ .}

\subsec{Brane bending and stabilization}

If we put a stress tensor of the form \sourceT\ on the brane, in general 
it will bend the brane; 
this phenomena was studied in \refs{\GaTa,\GKR} and
we will extend the analysis there to treat the present situation.
For example, we'll find that (as expected) a point mass placed on the
brane  will bend it further into the infrared.  

However, in typical examples of non-conformal theories, such as the
N=1* theory \refs{\PoSti}, the brane cannot be at an arbitrary
location, but its position is fixed by the perturbation away from
conformality. In
\PoSti, this stabilization results from non vanishing of components of
the G-flux.\foot{I thank J. Polchinski for a discussion on this
point.}    This has the effect of giving an effective potential for
the location of the brane, in other words a mass to the ``radion''
which describes shifting the position of the brane.
In general
this will be summarized in a stress tensor ${}^{\rm stab}T_{IJ}$ on
the RHS of Einstein's equations, \gyy-\munueq.
If the radion
is very massive, the brane will bend relatively little when the
vacuum solution is perturbed.  

In the $N=1^*$ theory, deformations of the radion are related to
fluctuations of the form $\langle tr \phi^2\rangle $.  
Since the dynamics of \PoSti\ are rather complicated, we will instead
use a toy model for radion stabilization, along the lines suggested in
Cs\'aki et.~al.\refs{\CGRT} and in \GKR, and which can be motivated by
studying properties of the Goldberger-Wise model for radion
stabilization\refs{\GoWi}.  Specifically, suppose that we
work in straight gauge, where $h_{yy} = h_{yy}(x)\neq0$; in this gauge,
fluctuations of the radion correspond to fluctuations of $h_{yy}$.  
A particularly simple form of the stabilizing stress tensor,
compatible with energy-momentum conservation, is
\eqn\stabstress{\eqalign{ {}^{\rm stab}T_y^y&= \mu^2 M_P^{d-1}h_{yy}(x)
e^{yd/R}\cr {}^{\rm stab}T_{\mu I}&=0\ }}
where $\mu$ is a constant that is proportional to the mass of the radion.
One can think of this as roughly arising from a ``potential'' of the
form $U=\mu^2 (h_{yy})^2$, as in \CGRT, or derive it in a limit of
the Goldberger-Wise wise model as is discussed in the appendix.  A basic
intuition behind the possibility to have a non-zero $T_{yy}$ with other
components vanishing is that a very stiff spring can exert a large force
with little change in its energy.

We will study two limiting cases of such stabilization; the first $\mu
R\ll1$, in which brane bending is the dominant effect, and the second
where $\mu R\gg1$, and brane bending is effectively eliminated.
Although it may be possible to physically achieve either limit in
more fundamental models, one expects that commonly the radion mass
will take a value $\mu\sim 1/R$, complicating the analysis.  In
subsequent sections we will see that a particularly important physical
question is whether the radion is the lightest massive mode, or
whether the first Kaluza-Klein mode of the graviton is lighter.

\newsec{Black holes on IR boundaries?}

We would like to study the properties of strongly-coupled
gravitational solutions in the vicinity of the IR boundary of the
space, and specifically search for black holes. The simplest case is a
solution with the symmetries of d-dimensional Schwarzschild, namely
SO(d-1) rotation symmetry.  We look for a solution with matter source
on the IR boundary or brane.  The long range weak field of such a solution
should correspond to the long range weak field of a point mass at the
boundary.  Thus, from the linearization principle, we should be able
to detect the presence of a full non-linear solution  and it's gross
features from studying the linearized equations.
Of course, for a given linear solution
there could be more than one non-linear solutions, or none;
to address the former, 
we assume that there is a uniqueness result analogous to Birkhoff's
theorem; if this is not true, then that has other potentially interesting
consequences.  And we should also bear in mind that 
including realistic dynamics for other fields
besides the radion and graviton could lead to other richer phenomena.

Therefore consider a point mass on the boundary, 
\eqn\ptmass{S_{\mu\nu}= 2m \delta^{d-1}(x) \delta_\mu^0\delta_\nu^0\ ,}
(the factor of two arises from the orbifold treatment of boundary
conditions) with a bulk stress
tensor that stabilizes the radion, \stabstress, but otherwise
vanishes.  In the next subsections we'll solve for the linearized
field of a general source on the brane, in the limits of 
stiff or soft radion, and then specialize to the case of such a point mass.

\subsec{The light radion limit}

In the case $\mu R\ll 1$ we neglect the stabilizing stress tensor
entirely.  This problem is most easily solved by passing to bent
gauge, with $h_{yy}=0$.  The divergence of the $(\mu y)$ Einstein 
eqn.~\muysol\ can then be
integrated with respect to $y$ to find
\eqn\ddhsol{\partial^\mu\partial^\nu h_{\mu\nu} = \partial_d^2 h +
f(x)\ ,}
for unknown $f(x)$.  Using this in the $(yy)$ equation \gyy\ then gives
\eqn\yysoln{{d-1\over R} \partial_y h = -f(x) e^{2y/R}\ .}
This is integrated subject to the boundary conditions \bcondb\ and
those at infinity.  Demanding that the field die off at infinity
requires that $f\equiv 0$, and therefore that $\partial_y h\equiv 0$.
The trace of \bcondb\ then determines the position of the boundary,
\eqn\Lsoln{\partial_d^2 L = {1\over 4(d-1)  M_P^{d-1}} S(x) \ .}
The $(\mu y)$ equation then implies 
\eqn\Muy{\partial_y\partial^\nu h_{\mu\nu} =0\ ,}
and the gauge freedom \dgauge\ can then be used to go to transverse gauge,
\eqn\transvg{\partial^\nu {\bar h}_{\mu\nu} =0\ .}
Thus the $(\mu\nu)$ equations reduce to 
\eqn\munured{\sq h_{\mu \nu} = 0\ .}
This is solved using the Neumann Green function for the scalar
laplacian,
defined by
\eqn\branegf{\eqalign{
\sq \Delta_{d+1} (X,X^{\prime}) &= 
{\delta^{d+1}(X - X^{\prime})\over {\sqrt {-
G}}}\ ,\cr
\partial_y \Delta_{d+1}(X,X')\vert_{y=0}&=0\  }}
with coordinates $X=(x,y)$ or $(x,z)$.  Using the boundary conditions
\bcondb, the solution is 
\eqn\bentsoln{h_{\mu\nu}(X)= - {1\over 2 M_P^{d-1}} \int d^d x' \sqrt{-g}
\Delta_{d+1}(X;0,x') \left[ S_{\mu\nu}(x') - \eta_{\mu\nu}
{S_\lambda^\lambda(x') \over d-1} + {\partial_\mu\partial_\nu \over
\partial^2} {S_\lambda^\lambda(x') \over d-1}\right]\ .}

Specializing to the point mass \ptmass, we first solve for the radion from
\Lsoln, using the $d$-dimensional Green function.  The result is
\eqn\Lresult{L= {m\over 2(d-1)M_P^{d-1}} {1\over (d-3)\Omega_{d-2} 
r^{d-3}}\ }
where $\Omega_n$ is the volume of the unit sphere $S^n$.
The metric can likewise be found using \branegf.  In particular, 
\eqn\httsol{ h_{00} = -{m\over M_P^{d-1}} {d-2\over d-1} \int dt'
\Delta_{d+1}(X;0,{\vec 0}, t')\ }
gives the effective scalar potential.

The shape of gravity  is thus determined by the scalar Green
function.  This is found in the appendix, and, for source on the brane,
simplifies to 
\eqn\scgf{\Delta_{p+1}(x,z; x,R) = -\left({z\over R}\right)^{d/2} \int
{d^dp \over (2\pi)^d} {1\over q} {J_{d\over2}(qz)\over J_{{d\over 2
}-1}(qR)} e^{ip(x-x')}\ .}
with $q^2=-p^2$.

While the expression \bentsoln\ is difficult to evaluate explicitly, we are
particularly interested in the long-distance limit where it simplifies.
Specifically, assume that $z\ll R$ and/or $x\gg R$.  In either case, the
integral is then dominated by the region of small $qz$, and so we can
replace the Bessel functions by a small argument expansion.  In particular,
in the case of the point mass \ptmass, we find
\eqn\hlong{ h_{00}(X)\simeq {mR\over (d-1) M^{d-1}} \left({z\over R}\right)^d
 \int {d^{d-1}p\over (2\pi)^{d-1}} {(qR/2)^{{d\over 2}-1} \over J_{{d\over
2}-1}(qR)} 
{d-2\over \Gamma({d\over 2}+1)} e^{i{\vec p}\cdot{\vec x}}
\ .}
The denominator has poles where $qR$ is one of the zeroes $j_{d/2-1,n}$ 
of the Bessel function
$J_{{d\over 2}-1}$.  Therefore we expect the large-$r$ behavior to be
dominated by the first pole, and specifically, the potential to take the
form
\eqn\potasy{ h_{00}\simeq {km\over  R M_P^{d-1}} \left({z\over
R}\right)^d {e^{-M_1 r}\over r^{d-3}}}
where $k$ is a numerical constant and 
the mass of the lightest Kaluza-Klein mode of the graviton is given by
\eqn\firstKK{M_1 = j_{{d\over 2}-1,1}/R\ .}

Note that, as expected, there is no $1/r^{d-3}$ falloff in $h_{00}$ 
that would be
characteristic of $d-$dimensional gravity.  Without an ultraviolet
brane, the effective $d-$dimensional Planck mass is infinity; in other
words, there is no graviton zero mode.

A sufficiently large point mass might be expected to produce 
a black hole.  Specifically, 
for a given mass, the horizon would be expected to lie in the 
region where $h_{00}\sim 1$, namely where $g_{00}$ is expected to
vanish, 
and 
for sufficiently large $m$ the size of this region would be
expected to be larger than $R$.  However, in the present case one
cannot explicitly find such a linearized solution.  From \Lresult, we find that the
condition for the radion field to stay linear, \smallb, is 
\eqn\linbb{r\roughly> \left({m\over M_P^{d-1}}\right)^{1\over d-2}\ .}
In other words, if we concentrate a mass $m$ in a region smaller than
that given by this equation, the brane bending goes non-linear and
forces us to do a non-linear analysis.  One can easily convince
oneself that in the present case this happens before $h_{00}\sim1$.
One might extend this analysis by solving the non-linear generalization of
\Lsoln\ and then find the metric perturbation satisfying \munured\ in this
background, but we leave this for further work.
While there are strong gravitational effects at radii less than
those given by \linbb, and while those may well correspond to black hole
formation, we can not presently clearly interpret them as such without
treating the physics of the large bending of the boundary.

\subsec{The heavy radion limit}

We next solve Einstein's equations \gyy-\munueq\ in the stiff limit,
$\mu R\gg 1$.  With stabilization present, we work in straight gauge.
Taking the divergence of the $(\mu y)$ equation \muysol\ and
integrating with respect to $y$ gives
\eqn\muysols{\partial^\mu\partial^\nu h_{\mu\nu}-\partial_d^2 h =
{d-1\over R} y \partial_d^2 h_{yy} + f(x)\ ,}
again with unknown integration constant $f(x)$.  Combining this with
the $(yy)$ equation \gyy\ then gives 
\eqn\yysols{\partial_y h = -y e^{2y/R} \partial_d^2 h_{yy} - {Rf\over
d-1} e^{2y/R} - {d\over R} h_{yy} - {\mu^2 R \over d-1} h_{yy}
e^{yd/R}\ .}
This must again be solved with boundary conditions \bcond\ and 
that the field fall
off sufficiently rapidly at infinity.  

In the limit of large $\mu$, the boundary condition \bcond\ at $y=0$ can be
solved by taking
\eqn\largmuh{ \mu^2 R h_{yy} = {S\over 2M_P^{d-1}}\ ,}
which has solution with $h_{yy}\sim {\cal O}(1/\mu^2)$: with a large
mass, the deformation of the radion is small.  This is precisely as in
the cosmological case in \CGRT.  The boundary condition at infinity
then implies that $f=0$. In the stiff limit $\mu\rightarrow \infty$,
we therefore simply set $h_{yy}=0$, and solve the remaining equations using
\eqn\DHDY{\partial_y h = {e^{dy/R} S\over 2(1-d) M_P^{d-1}}\ .}

Indeed, returning to the $(\mu y)$ eqn., we find that up to a piece
that can be gauged away by \dgauge, 
\eqn\tranvpart{\partial^\nu {\bar h}_{\mu\nu} = {R\over 4 d(1-d)
M_P^{d-1}} e^{dy/R} \partial_\mu S\ .}
The remaining $(\mu \nu)$ equation then becomes
\eqn\munusols{\sq {\bar h}_{\mu\nu} = {R\over 2d(1-d) M_P^{d-1}}
e^{(d+2)y/R} \left(\partial_\mu\partial_\nu S- \eta_{\mu\nu}
{\partial_d^2 S\over 2} \right)\ .}

This final equation is most easily solved by using a
transverse-traceless projection; specifically, define the transverse
projection operator
\eqn\tproj{\Pi_{\mu\nu} = \eta_{\mu\nu} - {\partial_\mu \partial_\nu
\over \partial_d^2}\ .}
Then the transverse-traceless piece of ${h}$ is given by
\eqn\hTT{ {h}_{\mu\nu}^{TT} = \Pi_{\mu\lambda} \Pi_{\nu\sigma}
{\bar h}^{\lambda\sigma} -{1\over d-1} \Pi_{\mu\nu} \Pi_{\lambda\sigma}
{\bar h}^{\lambda \sigma}\ }
and is easily shown to satisfy
\eqn\TTeqn{ \sq {h}_{\mu\nu}^{TT} = 0\ .}
The boundary condition is likewise projected and gives
\eqn\ttbc{\partial_y{h}_{\mu\nu}^{TT}
\vert_0   = {1\over 2M_P^{d-1}}\left( S_{\mu\nu} -
{1\over d-1} \Pi_{\mu\nu} S\right)\equiv {S_{\mu\nu}^{TT}\over 
2 M_P^{d-1}}\ .}
These are again solved using the scalar Green function, and give
\eqn\TTsoln{ {h}_{\mu\nu}^{TT} = -{1\over 2M_P^{d-1}} 
\int d^d x' \sqrt{-g}
\Delta_{d+1}(X;0,x') S_{\mu\nu}^{TT}\ ,}
exactly as in \bentsoln. 
The full metric perturbation is then easily found to be
\eqn\hsols{h_{\mu \nu} = {h}_{\mu\nu}^{TT} + {\partial_\mu
\partial_\nu\over \partial_d^2} h\ }
with $h$ given by the integral of \DHDY.  In particular, the linearized
potential of a point-like static source on the brane is once again given by
\httsol. 

\ifig{\Fig\figone}{Shown is a sketch of the strong gravity region in the 
rigid brane limit. This region should be smoothed out in a summit 
region of size $\Delta x\sim {\cal O}(R)$.  This curve should give
the approximate shape of a black hole horizon.}{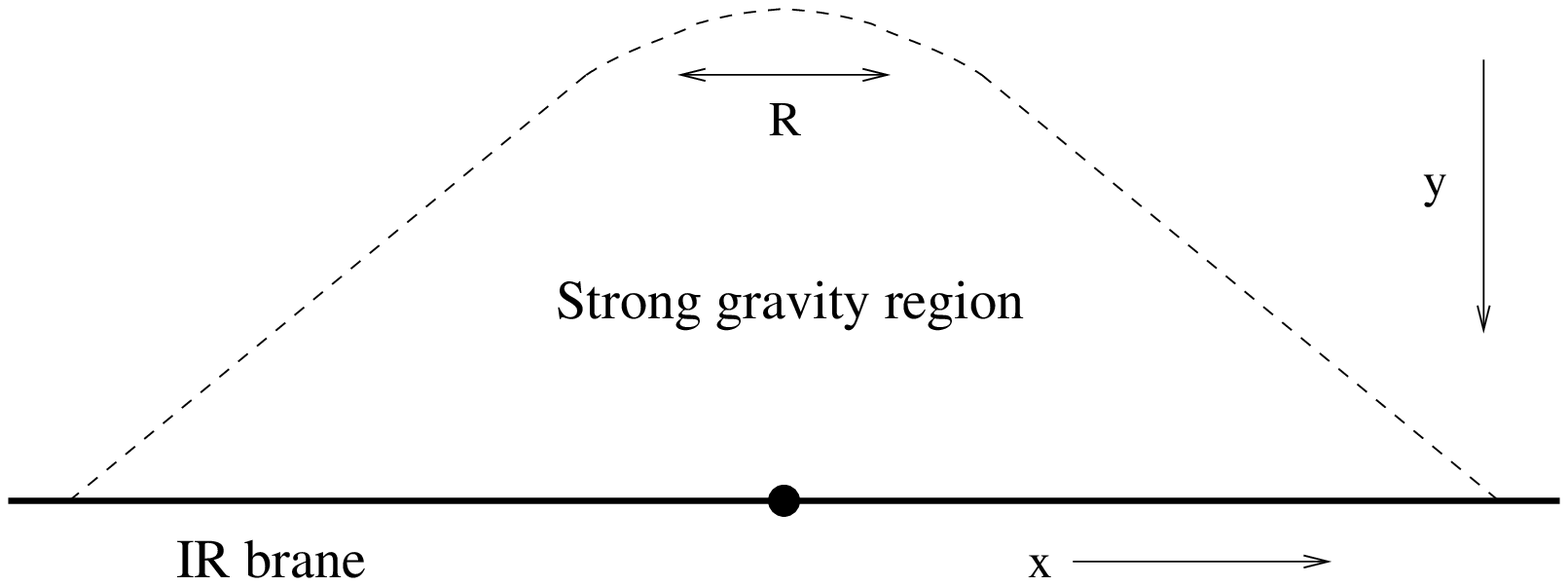}{2.0}

We can again inquire regarding the existence of a black hole solution.
Asymptotically far from the source, $h_{00}$ is again given by the simple
form \potasy.  As
we approach the source, $h_{00} \approx 1$ at a surface given by 
\eqn\horizeq{M_1 r-yd/R \simeq  \ln\left[{km\over R 
M_P^{d-1}r^{d-3}}\right]\ .}
This surface is sketched in figure 1.  We expect that in the exact solution
it is rounded off in a region of size $\sim R$ surrounding the summit.  
Its size grows logarithmically with
the mass $M$.  By the linearization principle, this should be the shape of
the corresponding black hole horizon.  Note that the size of the region,
projected onto the infrared boundary, is given by
\eqn\horsize{r_h(m)\sim {1\over M_1} \ln \left[{km M_1^{d-3}\over R 
M_P^{d-1}}\right]\ .}

As pointed out by L. Susskind\Suss, there is a question about the 
existence of stable black
holes with this shape; he in particular has argued that any such horizon
would thermodynamically prefer to sink into the boundary.  To make this
more precise, a quick estimate
shows that there could be higher entropy solutions with the same conserved
energy. 
This is most easily studied if we imagine that we periodically
identify the flat dimensions so that they have finite volume $V$; for
simplicity consider the case of $d=4$.  The
entropy or area of the plateau at the summit goes like 
\eqn\entest{A\sim \left(e^{-{y_{\rm plateau}\over R}} R\right)^3 \sim R^3 
\left[{km\over R^2 M_P^3}\right]^{3/4}\ .}
On the other hand, we might guess that another configuration with the same
energy is a black brane, with an energy density $\epsilon \sim m/V$. We can
estimate the entropy as that of the usual black three-brane solution, see \eg\
\refs{\GuKP}: 
\eqn\bbent{S\sim \sqrt{N} V^{1/4} m^{3/4}\ .}
In the large volume limit, the black brane entropy is clearly larger.

However, note one critical caveat.  Our linearized analysis assumed the
existence of a stabilization mechanism.  For the present discussion of the
instability to decay to a black brane to be relevant, there must be a black
brane solution to the coupled equations including the stabilizing fields.
The question of the existence of such solutions -- which corresponds to the
question of finding a finite temperature vacuum of the field theory --
has been the subject of some discussion; see \eg\
\refs{\Gubs}.  In some cases it is believed that the field theory supports
no corresponding thermally excited state. However, one may check that 
there is a limit of the Goldberger-Wise mechanism as $m\rightarrow0$ 
in which the radion mass stays fixed but the backreaction of the 
Goldberger-Wise field at zero radion displacement vanishes.  This 
suggests that such simple models have black brane solutions like 
those where there is no stabilization mechanism present.   The existence 
of a stable black brane solution in a general scenario is an 
interesting and
important question that will ultimately determine the fate of the solutions
found in this paper, and different theories may yield different results.

\subsec{Intermediate radion mass}

We now give a general discussion of the case where the radion mass falls
between the two limiting cases.  In this case it is more difficult to solve
the corresponding coupled equations, but we hope to infer basic properties
of the solutions.\foot{Note, however, that by a general argument (see
\refs{\DFGK} eqn.~(62)) it is always possible to solve for the transverse
traceless part of the metric given the scalar Green function.  Solving for
the coupled longitudinal, trace, and radion excitations is more complicated.}

In this intermediate case, it may be most advantageous  to again work in
bent gauge.  We can guess the contribution of the stabilizing stress tensor
to the equations of motion in this gauge; in particular, we expect the
boundary condition and equations of motion to give
\eqn\newBC{ \partial_d^2L + M_L^2 L \sim {S(x)\over M_P^{d-1}}\ }
where $M_L\propto \mu$ is the mass of the radion.
Outside a spherical mass distribution with large mass $m$, we then expect the
radion to take the form
\eqn\Lsoln{ L\sim {m\over M_P^{d-1}} {e^{-M_L r}\over r^{d-3}}\ .}

Now suppose that we take such a mass distribution and gradually compress
it; we'd like to know what becomes important first, large bending, or
strong gravity/possible horizon formation, $h_{\mu\nu}\sim \calo( 1)$.
Obviously this depends on the relative magnitude of the radion mass to the
mass of the first KK excitation of the graviton.  Bending becomes important
when the mass is concentrated in a region with size given by
\eqn\bendimp{ {mM_L\over M_P^{d-1}} {e^{-M_L r}\over r^{d-3}}=  \calo(1)\ .}
Other strong gravity effects such as a horizon would be found where
\eqn\bhimp{ {m\over R M_P^{d-1}} {e^{-M_1 r}\over r^{d-3}} = \calo(1)\ .}
For large $m$, strong gravity becomes important first if 
$M_L>M_1$, and bending goes
nonlinear first if $M_L<M_1$.  From the perspective of the dual field
theory, this is obvious:  the lightest field dominates the dynamics at long
distances.

\subsec{Black hole formation in TeV-scale gravity}

In the context of TeV-scale gravity, 
recent work \refs{\GiTh,\DiLa} 
has examined black hole formation and decay
in high energy collisions.  In the most optimistic of scenarios such
processes could be visible at LHC.  Once one pushes to energies higher than
the fundamental 
Planck scale, one makes bigger and bigger black holes, and the
properties of these black holes might be used to infer aspects of the
geometry of the extra dimensions \refs{\GiTh}.  One possibility is that the
extra dimensions are large and approximately flat, in which case the black
hole production cross section at energy $E$ goes like
\eqn\bhflat{\sigma\sim E ^{2\over D(E)-3}}
where $D(E)$ is the number of spacetime
dimensions large compared to the Schwarzschild
radius of the black hole at the given energy.  Another possibility is that,
at large enough scales, the extra dimensions in the vicinity of our
observable brane take the form of a piece of AdS, like in the toy model of
\refs{\RaSui} or the string solutions of \refs{\GKP}.  In that case, once
one reaches energies large enough to make black holes bigger than the AdS
radius scale, the cross section is given by \horsize, and grows
logarithmically:
\eqn\crosssize{\sigma\sim R^2 \ln^2 \left[{ER\over (M_P
R)^{d-1}}\right]\ .}
Of course, the above comments and questions regarding possible dominance of
radion excitations, classical stability of the resulting black holes,
\etc\ would have very important phenomenological consequences.  It would be
particularly interesting to better understand any instability to black brane
formation more deeply.  However, closer investigation of these questions 
likely requires more detailed constructions.

\newsec{High energy QCD scattering and the Froissart bound}

We now return to investigate aspects of the dual gauge theory description
of this physics.  Corresponding to different solutions, there are various
realizations of dual gauge theories, but there are some generic features.
We expect these to include the presence of the four-dimensional field that
corresponds to the radion, as well as the tower of KK excitations of the
graviton that we think of as glueballs.  In addition there may be other
fields.  The subsequent discussion applies in cases where these other
fields do not significantly alter the dynamics.

Polchinski and Strassler\refs{\PoSt} investigated the role of bulk physics
in producing parton-like behavior.  However, at large energies we don't
expect this to be the dominant physics.  From the bulk point of view this
is clear: at large energies the cross section for strong gravitational
effects, such as black holes with sizes given by \horsize, grows with
energy and is expected to be a dominant effect in the physics.
Indeed, in QCD it is also known that the total cross-section is not
dominated by hard processes, and grows with energy.  Let us compare
these more closely.

As described in subsection 4.3, there are two possible 
explanations for the dominant
long-distance non-linear gravitational behavior: brane bending, or strong
gravity effects, such as black hole formation, mediated by the first Kaluza
Klein mode of the graviton. Which is dominant depends on the relative
magnitude of their masses, $\mu$ and $M_1$.  However, in either case, at
high center of mass energy, the size of the regions where they become 
important follows
a simple scaling law. Radion dominant scattering processes should set in at 
\eqn\raddom{r_L\approx {1\over M_L} \ln \left( {M_L^{d-2} E\over
M_P^{d-1}}\right) - {d-3\over M_L} \ln\ln\left({M_L^{d-2} E\over
M_P^{d-1}}\right) + \cdots\ .}
KK graviton dominant processes should set in at $r\sim r_h$ as given in
eqn.~\horsize, or equivalently \raddom\ with $M_L$ replaced by $M_1$.  
In either case one can estimate the size of the scattering
cross-section,
\eqn\crossest{\sigma\sim \pi r_{\rm L\, or\, h}^2\ ,}
which gives\foot{Note that \APR\ also considered this problem, but came to
the different conclusion that the cross-section ceases to grow after black
holes reach size $\calo(R)$.} 
\eqn\crossestii{\sigma \sim {\pi\over M_L^2} \ln^2 \left( {M_L^{d-2} E\over
M_P^{d-1}}\right)\ \ {\rm or} \ \ \sigma \sim {\pi\over M_1^2} \ln^2 
\left({EM_1^{d-2}\over M_P^{d-1}}\right)\ .}
In either case we have recovered from bulk physics 
the $\log^2E$ behavior associated with the
Froissart bound, which is a general bound following from unitarity!\foot{For a 
recent review and discussion of the Froissart bound, see \refs{\Levin}.}
To date there has been no solid argument that 
QCD saturates this
bound at high energies, although this likelihood has been widely discussed.

In hindsight, reproducing this bound is not so surprising as it may seem, and
in fact is related to an old heuristic argument for saturation
given by Heisenberg in 1952\refs{\Heis}.  He argued that a target hadron is
surrounded by a pion field with energy density $\sim e^{-m_\pi r}$, and
suggested that inelastic processes will occur when the collision is close
enough to locally yield enough energy to create a pion pair.  This yields
an estimate
\eqn\Heis{\sigma \sim {1\over m_\pi^2} \ln^2{E\over m_\pi^2}\ ,}
similar to the above.

It is amusing to comment that the present discussion suggests that, via the
AdS/CFT duality, the bulk physics ``knows'' about boundary unitarity, and
that furthermore, the high energy scattering is approximately described by
classical bulk physics.

Note also that, in addition to strong gravitational scattering/black hole
formation in the vicinity of the IR boundary, as the energy increases,
formation of black holes above the IR boundary becomes possible.  However,
for a given gauge theory energy, the local energy is largest in the
vicinity of the brane, and given that the gravitational cross section rises
with energy, it seems plausible that the dominant processes are those
localized near the IR boundary.

Of course, we'd like to understand more features of the scattering
physics.  In the radion-dominated case, this question depends closely on
the dynamics of the radion, whose investigation we leave for future work.
In the KK graviton-dominated case, the outcome depends intimately on the
stability of non-linear black hole solutions that would correspond to the
linearized solutions found in the preceding section.  There are two
possibilities.  One is the classical instability to decay into a black
brane.  L. Susskind has suggested that in this case the gravitational
solution would sink into the IR boundary and spread out, ultimately forming
a black brane at infinitesimal temperature.  The corresponding gauge theory
dynamics might be described as formation of a fireball that gradually cools
as it spreads out and thermalizes.  One expects relevant time scales to be
given by $t\sim r_h$.  Alternatively, if there is no classical instability,
one must consider quantum processes.  One is tunneling to a black brane, if
one exists.  If this process is sufficiently slow, or if a corresponding
black brane solution does not exist, then plausibly the dominant decay is
via Hawking evaporation of the black hole.  Note that this is expected to
have a time scale parametrically larger in the collision energy.

\newsec{Conclusions and outlook}

Understanding high-energy scattering behavior in gauge theories has
remained a historically challenging problem.  This suggests turning the
lens of AdS/CFT duality to focus on it.  This paper has argued that a
generic phenomenon on the gravitational side, strong gravitational physics
such as black hole formation, could play a central role in the high energy
physics of the gauge theory.  Furthermore, black hole formation at high
energies is a nearly classical process:  for high center of mass energies,
horizons can 
form at weak curvature\refs{\EaGi}, rendering quantum effects
sub-dominant.  

Unfortunately we aren't yet able to describe the non-linear gravitational
solutions relevant to this physics, and to do so probably also requires
better knowledge of the bulk string theory backgrounds, such as that in
\PoSti, in which these processes take place.  Nonetheless, this paper has
presented, in an ``effective theory'' approach, an analysis of the linear
behavior of the gravitational field.  The linear approximation also
contains information about it's demise, and thus indicates in what regions
non-linear effects are important.  In particular, using this approximation,
one may estimate high energy scattering cross sections for gauge theory
processes, and derive the $\sigma\sim \ln^2 E$ behavior saturating the
Froissart bound.  This suggests that bulk physics in a sense ``knows''
about boundary unitarity -- though it is not clear what implications, if
any, this has for the problem of bulk unitarity.

This picture also suggests a means to investigate the subsequent
evolution of the scattered state, although to do so in detail seems to
again require more knowledge of the string backgrounds and the non-linear
solutions that live in them, and in particular of the stability properties
thereof.  In one picture a black hole, once formed, ``melts'' onto the
infrared boundary.  In other configurations, there may even be quasi-stable
black holes (only destabilized by Hawking evaporation); it is very interesting to
contemplate properties of 
the dual description of such an object in the gauge theory.

Open questions therefore include more thorough examination of these processes in more
detailed models.  In particular, one may start with the gravity dual of the
$N=1^*$ theory or other analogues.  As pointed out, two of the relevant
features of the dynamics are the existence of gravity in the bulk, and of a
stabilization mechanism for the position of the infrared boundary, that is
the radion field.  It would be interesting to better understand the
relation between the stabilization of \PoSti\ and the Goldberger-Wise
mechanism\GoWi, and the question of whether there are more interesting
effects beyond those summarized by the effective stabilization stress
tensor \stabstress.  We would like to know what possibilities exist for the
radion mass:  are there indeed consistent scenarios with the radion mass
exceeding the mass of the first Kaluza-Klein mode, so that analogs to black
holes can be relevant, and is a large radion mass (as compared to $1/R$)
possible?  We would like to understand the properties of the resulting
non-linear gravitational solutions, and to understand whether they are
classically unstable.  The latter hinges on the question of the existence
of black brane solutions in the presence of the other fields that provide
stabilization and other dynamics; the corresponding question in the gauge
theory is whether the theory supports a finite temperature phase.  
It is certainly conceivable that a
variety of models result in a variety of different effects.  

This paper has
probably just scratched the surface of the role of strong gravitational
effects in their Maldacena dual gauge theories.

\bigskip\bigskip\centerline{{\bf Acknowledgments}}\nobreak

I'd like to thank T. Degrand, O. DeWolfe, D. Gross, S. Kachru, 
J. Polchinski, 
and M. Strassler for very valuable conversations, and in particular
N. Arkani-Hamed, L. Susskind, 
and S. Thomas for seminal discussions.
This work was
supported in part by the Department of Energy under
Contract DE-FG-03-91ER40618, and by David and Lucile Packard 
Foundation Fellowship 2000-13856.

\appendix{A}{Scalar Green function for AdS with an infrared brane}

This appendix derives the scalar Neumann Green function in an anti-de
Sitter space that has been truncated in the infrared.  More general
supergravity backgrounds, where the backreaction of other fields are
important, will entail more complicated generalizations of this correlator.

The Neumann Green function is defined to be the solution of the equation
\eqn\braneGF{
\sq \Delta_{d+1} (X,X^{\prime}) = 
{\delta^{d+1}(X - X^{\prime})\over {\sqrt {-
G}}}\ }
where $X=(z,x)$ or $(y,x)$, and the boundary condition is 
\eqn\Neubc{
\partial_y \Delta_{d+1}(X,X')\vert_{y=0}=0\  .}

This may be solved by the ``method of matching,'' as in \GKR.
Specifically, the Green function solves the homogeneous Laplace equation
for $z>z'$ and for $z<z'$; these two solutions may then be matched by
integrating \braneGF\ across the singularity at $z=z'$.  In order to do so,
we work with the Fourier transform,
\eqn\ftgf{ \Delta_{d+1}(X,X') = \int {d^d x\over (2\pi)^d} e^{ip(x-x')}
\Delta_p(z,z')\ .}
Solutions to Laplace's equation come in the form of superpositions of 
Bessel functions,
$z^{d/2}J_{d/2}(qz)$, $z^{d/2}Y_{d/2}(qz)$, with $q^2=-p^2$.  
We write the general superposition of such solutions for $z>z'$ and $z<z'$,
apply the boundary condition \Neubc, demand that the Green function be
regular in the UV at $z=0$, and match by integrating \braneGF\ from
$z=z'-\epsilon$ to $z=z'+\epsilon$.   The result is straightforwardly found
to be 
\eqn\scgfs{\Delta_p = {\pi \over 2 R^{d-1}} {(zz')^{d\over 2}\over
J_{{d\over 2} -1}(qR)} \left[ Y_{d\over 2}(qz_>) J_{{d\over 2}-1}(qR) -
J_{d\over 2}(qz_>) Y_{{d\over 2}-1}(qR)\right] J_{d\over 2}(qz_<)\ .}

The Green function \ftgf\ significantly 
simplifies with one point on the brane:
\eqn\gfbrane{ \Delta_{p+1}(x,z; x,R) = -\left({z\over R}\right)^{d/2} \int
{d^dp \over (2\pi)^d} {1\over q} {J_{d\over2}(qz)\over J_{{d\over 2
}-1}(qR)} e^{ip(x-x')}\ .}

We also need the asymptotics of this Green function.  At either large $x$
or small $z$, dominant contributions come from the region with
$qz\ll1$. This means that we can make a small argument expansion in $qz$ to
find
\eqn\Deltaasy{\Delta_{d+1}(X;0,x') \simeq -{1\over \Gamma({d\over 2}+1)}
\left({z\over R}\right)^{d\over 2} \int {d^d p\over (2\pi)^d } {1\over q}
\left({qz\over 2}\right)^{d\over 2} {e^{ip(x-x')}\over J_{{d\over
2}-1}(qR)}\ .}

In particular, note that the integrand in \Deltaasy\ has no pole at $q=0$
-- as expected, in this limit the volume is infinite and there is no
massless d-dimensional graviton.  It does have poles at the masses of each
of the Kaluza-Klein modes of the graviton.  
At long distances the integral is
dominated by the mass of the lowest Kaluza-Klein mode, corresponding to the
first zero of the Bessel function:
\eqn\kkmass{M_1={j_{{d\over 2}-1,1}\over R}\ .}
%
The static Green function is gotten by integrating \ftgf\ over time.  Its
asymptotic behavior is given in terms of this mass:
\eqn\AsyGF{\int dt' \Delta_{d+1}(X;0,{\vec0},t')\simeq {\hat k}
\left({z\over R}\right)^d {e^{-M_1r}\over R r^{d-3}} }
where ${\hat k}$ is a numerical constant.

\appendix{B}{Stabilization, Goldberger-Wise or otherwise}

In this appendix we outline the basics of the Goldberger-Wise mechanism\GoWi,
with particular emphasis on large (in the limit, infinite) brane
separation, and relate it to the stabilization stress tensor \stabstress\
used in the text.  

We begin by thinking of the situation with AdS truncated by a brane in the
IR and one in the UV.  
The basic idea of the Goldberger-Wise mechanism is to postulate the
existence of a massive bulk field, with lagrangian
\eqn\GWlag{S[\phi]= -\hf \int d^{d+1} X \sqrt{-G} \left[ (\nabla \phi)^2 +
m^2 \phi^2\right]}
and such that the {\it value} of the field is fixed both on the IR and UV
branes.\foot{Goldberger and Wise actually considered more generally a
potential for the values of $\phi$ on the boundary, but we omit this
unneeded generalization.}  Bringing the branes either too close or too far
raises the value of the action $S[{\bar \phi}]$ at the corresponding 
static solution ${\bar \phi}$ of the equations of motion.  This therefore
generates a potential for the radion.

Note that the Goldberger and Wise's original analysis assumed $m^2>0$, but
the mechanism works for $m^2<0$ \DFGK.  This is presumably related to 
the stabilization evident in Polchinski and Strassler's
work, which involves a CFT perturbation that is relevant, hence has
dimension $\Delta<4$, corresponding to $m^2<0$.  While we expect the
dynamics to be qualitatively similar, it would be interesting to further
elucidate the precise relationship between the two schemes.

We are interested in applying the mechanism to the limiting case where the
UV brane is moved to infinity.  We therefore generalize the boundary
condition at this end of the space to state that as $y\rightarrow-\infty$,
\eqn\asybeh{\phi\rightarrow \phi_0(e^{k_-y} + A e^{k_+ y})}
for fixed $\phi_0$ and arbitrary $A$.  Here the exponents are the familiar quantities
\eqn\kpmdef{k_\pm= 
{d\over 2R} \pm \sqrt{{d^2\over 4R^2} + m^2}\ .}
The boundary condition at $y=L$ is then taken to be $\phi(L)=\phi_L$,
fixing $A$.  

If we compute the action \GWlag\ as a function of $L$, we find a potential 
\eqn\stabPot{\eqalign{V &= \hf\int dy e^{-dy/R}
\left[(\partial_y {\bar \phi})^2 + m^2
{\bar \phi}^2 \right]\cr &= {\phi_0^2\over 2} \left[
{d\over R} e^{(k_--k_+)L} - 2k_+ {\phi_L\over \phi_0} e^{-k_+
L} + k_+ \left({\phi_L\over \phi_0}\right)^2 e^{-(k_++k_-)L} + C\right]\ ,}}
where $C$ is a cutoff dependent but $L$ independent term.  This potential
has extrema at 
\eqn\Lmin{e^{k_-L} = {\phi_L\over \phi_0} {k_+^2 \pm \sqrt{k_+k_-(k_+k_- -
k_+^2 + k_-^2)}\over {k_+^2 -k_-^2}}\ .}
The radion mass, given by $V^{\prime\prime}(L)$, grows with the mass $m$ of
the scalar field; thus the limit of large scalar mass is one way of
motivating the
limit of large radion mass used in section 4.2.

Note we can also verify the approximate form of the stress tensor used in
\stabstress.  From \GWlag\ we find
\eqn\strestens{T_y^y\propto (\partial_y {\bar \phi})^2 -m^2 {\bar \phi}^2\ .}
The $L=0$ vacuum value of this is treated as part of the background solution.
The stabilizing  stress arises from the variation of this as $L$, thus
${\bar \phi}$, varies. In particular, we find 
\eqn\varT{{\partial T_y^y\over \partial L} \propto {\partial
A\over \partial L} \left[ -4m^2 + 2 A_0 (k_+^2 - m^2) e^{(k_+-k_-)y}
\right] e^{yd/R}\ .}
The first term is has the expected form of \stabstress.  The second term is
a correction that has support near the lower boundary at $y=0$, but
otherwise is small, and vanishes in the limit $m\rightarrow \infty$.  It also
may be checked that there is a limit $m\rightarrow \infty$, 
$\phi_0\rightarrow0$ such that the radion mass stays fixed (or goes to 
infinity), but the vacuum backreaction of $T_{IJ}(L=0)$ on the metric stays 
small.

\listrefs
\end

In particular, let's consider the idealized limit where the stabilization
mechanism is very ``stiff;''  this is the limit considered in
\refs{\CGRT}.  In other words, we imagine that the potential $U[h]$ has a
sharp minimum that fixes the IR boundary at a definite height, and that
there is an enormous cost of energy to move the boundary from this height.
(Deviations from this idealization in more realistic models such as \PoSti\
would be interesting to investigate.)
In particular, in this limit, the vertical position of the brane should be
a constant function of $x$, and so by a gauge transformation $y\rightarrow y'(y)$
we should be able to take $h_{yy}$ to be constant.  As in \CGRT, $U$ should
contribute a term of the form 
\eqn\stresstab{T_y^{y\,{\rm stab}}\propto U(h_{yy})+ h_{yy} U'(h_{yy})}
to the RHS of \gyy.  In the limit of infinitely stiff potential, this
equation is therefore solved by taking $h_{yy} = h_{yy}^0$, the value at
the minimum of the potential.  Without loss of generality, we can absorb 
$U(h_{yy}^0)$ into the definition of the cosmological constant, and use an
overall choice of scale $y\rightarrow ky$ to set $h_{yy}^0=0$.
In this limit eqns.~\gyy-\munueq\ simplify substantially.  In particular,
\muysol\ can be integrated to give
\eqn\muysol{ \partial^\nu h_{\mu\nu} = {1\over M_P^{d-1}} \int_0^y T_\mu^y\ ,}
and then eqn.~\munueq\ can be directly solved given the Green function
function for the scalar laplacian $\sq$.